\documentclass[prl,twocolumn]{revtex4}
\usepackage[english]{babel}

\usepackage{amsmath}
\usepackage{mathrsfs}
\usepackage{graphicx}

\def\>{\rangle}
\def\<{\langle}

\begin{document}
\title{Shape invariance in phase space}
\author{Constantin Rasinariu}
\affiliation{Columbia College Chicago, Department of Science and Mathematics\\ Chicago, IL 60605, Email: crasinariu@colum.edu}

\begin{abstract}
Shape invariance is a powerful solvability condition, that allows for 
complete knowledge of the energy spectrum, and eigenfunctions of a 
system. After a short introduction into the deformation quantization 
formalism, this paper explores the implications of the supersymmetric 
quantum mechanics and shape invariance techniques to the phase space 
formalism. We show that shape invariance induces a new set of relations 
between the Wigner functions of the system, that allows for their 
direct calculation, once we know one of them. The simple harmonic 
oscillator and the Morse potential are solved as examples.\\

\end{abstract}

\maketitle

\section{Introduction}

The phase space quantum mechanics formalism (also known as deformation
quantization) uses complex functions acting on the phase space instead
of operators acting on a Hilbert space. These functions are endowed
with a novel multiplication rule, the $\ast$-product, which is
non-commutative, associative, and hermitian. The formalism maps
operators $\hat A$ of quantum mechanics to complex functions $A(x,p)$,
and vice versa. Historically the fourth way of doing quantum
mechanics, the deformation quantization is an equivalent way of doing quantum mechanics.  As Dirac noted \cite{Dirac-51}
\begin{quote}
  Two points of view may be mathematically equivalent and you may
  think for that reason if you understand one of them you need not
  bother about the other and can neglect it. But it may be that one
  point of view may suggest a future development which another point
  does not suggest, and although in their present state the two points
  of view are equivalent they may lead to different possibilities for
  the future. Therefore, I think that we cannot afford to neglect any
  possible point of view for looking at Quantum Mechanics and in
  particular its relation to Classical Mechanics. 
\end{quote}
The goal of this work is to give a short introduction to the phase space 
quantum formalism, and to explore some of its implications as seen 
through the prism of the supersymmetric quantum mechanics (SUSQM). We 
show that shape invariance induces a new set of relations between 
the Wigner functions of the system, that allows for their direct 
calculation, once we know one of them. The paper is organized in three 
parts. In the first part we briefly present the main concepts of the 
phase space quantum mechanics formalism. In the second part we introduce 
SUSYQM and emphasize the role of shape invariance (SI) in the solvability 
of the system. In the third part we analyze some of the implications of 
SUSYQM and SI in deformation quantization and derive a recursion formula between the Wigner functions of the system. We calculate the simple 
harmonic oscillator and the Morse potential as a concrete examples.

\section{Quantum mechanics on phase space}

In the following, we will use the ``hat'' notation to symbolize quantum 
operators and the ``non-hat'' notation for the corresponding functions on the 
phase space. We will also consider only the one-dimensional case, where the 
phase space is simply $(x,p)$. For the general case, see \cite{Weyl-27}.

\subsection{The star product}
To any quantum operator $\hat A$ we associate a phase space function
$A(x,p)$ using the Weyl's transform $\mathscr{W}(\hat A)$
\cite{Weyl-27}, which in the coordinate $|x\rangle$ basis, reads
\begin{equation}
\label{weyl}
\mathscr{W}(\hat A) \equiv A(x,p)= \hbar \int dy \,e^{-ipy}\langle x + \frac{\hbar y}{2}\,|\, \hat A 
        \,|\,x - \frac{\hbar y}{2} \rangle ~.
\end{equation}
All  integrals run from $-\infty$ to $+\infty$, unless specifically 
restricted. The complex function $A(x,p) = \mathscr{W}(\hat A)$ is also known 
as the \emph{Weyl's symbol} of $\hat A$. Without other additional 
constraints, the reciprocal map $\mathscr {W}^{-1}$ is not unique. A 
sufficient condition for the unicity of $\mathscr{W}^{-1}$ is obtained by 
choosing an operator ordering \cite{Tosiek-95}. We choose Weyl's ordering, which 
prescribes symmetrical ordered polynomials in $\hat x$ and $\hat p$. I.e., $
\mathscr{W}^{-1}(x p)= (\hat x\hat p + \hat p \hat x)/2$.  Thus, we write the 
reciprocal map as
\begin{eqnarray}
\hat A (\hat x, \hat p) &\equiv& \mathscr{W}^{-1}(A(x,p)) \label{weyl-1} \\
&=&\frac{1}{(2\pi)^2}\int du\, dv\, dx\, dp\, A(x,p) \,
e^{i(\hat p - p)u + i(\hat x -x) v}\,.\nonumber
\end{eqnarray}
With this definition, $\mathscr{W}^{-1}(x)=\hat x$ and
$\mathscr{W}^{-1}(p)=\hat p$. More generally, one can readily check
that $\mathscr{W}^{-1}((a x + b p)^n) = (a \hat x + b \hat p)^n$.

Next, we define the phase space product between the Weyl's
symbols. This product should correspond to the product of the quantum
operators in the Hilbert space. Groenewold \cite{Groenewold-46} and Moyal
\cite{Moyal-49} showed that using Weyl's ordering assumption, the product
of functions on phase space can be written as
\begin{equation}
A(x,p) \ast B(x,p) = A(x,p)\,
  e^{\frac{i\hbar}{2}\left(
     \overleftarrow{\partial_x} \, \overrightarrow{\partial_p} 
   - \overleftarrow{\partial_p} \, \overrightarrow{\partial_x}
  \right) }
  B(x,p)\,,
\end{equation}
where the arrow indicates the direction in which the derivative
acts. Using (\ref{weyl}) one can prove that indeed $\mathscr{W}(\hat A
\hat B) = A(x,p) \ast B(x,p)$.

The Groenewold-Moyal product is also known as the star-product
($\ast$-product) of $A(x,p)$ and $B(x,p)$. It can be expressed in
several equivalent forms, such as
\begin{equation}
\label{star1}
A(x,p) \ast B(x,p) =  
   e^{
       \frac{i\hbar}{2} \left(
     \partial_x\, \partial_{p'} - \partial_p\,\partial_{x'}       
     \right)
     } A(x,p) B(x',p')\,,
\end{equation}
calculated at $(x',p')=(x,p)$; or, using the Bopp shifts \cite{Bopp-61} 
\begin{eqnarray}
\label{bopp}
A(x,p) \ast B(x,p) &=&
A(x+\frac{i\hbar}{2}\,\overrightarrow{\partial_p}\,,\,p-\frac{i\hbar}{2}\,
\overrightarrow{\partial_x})\,B(x,p) \\
&=& A(x,p)\,
B(x-\frac{i\hbar}{2}\,\overleftarrow{\partial_p}\,,\,p+\frac{i\hbar}{2}\,
\overleftarrow{\partial_x})\,. \nonumber
\end{eqnarray}
The $\ast$-product can also be represented as an integral, such as the
Fourier representation \cite{Neumann-31, Baker-58}
\begin{eqnarray}
\label{star2}
A(x,p) \ast B(x,p) &=& \frac{1}{(\pi\hbar)^2} 
   \int dx_1 dp_1 dx_2 dp_2 \\
   && A(x_1,p_1) B(x_2,p_2)   \nonumber \\
   && \times e^{ -\frac{2i}{\hbar}
   \left[
      \,p(x_1-x_2) + x(p_2-p_1) + (x_2 p_1 - x_1 p_2)
   \right]
   }\,,\nonumber
\end{eqnarray}
or as the alternate integral \cite{Hillery-83}
\begin{eqnarray}
\label{star3}
A(x,p) \ast B(x,p) &=& \frac{1}{(\pi\hbar)^2} 
   \int dx_1 dp_1 dx_2 dp_2 \nonumber \\
   && A(x+x_1,p+p_1) B(x+x_2,p+p_2) \nonumber  \\
   && \times e^{ \frac{2i}{\hbar}
   \left(
      \,x_1 p_2 - x_2 p_1
   \right)
   }\,.
\end{eqnarray}
We conclude this definition by emphasizing the cyclic, trace-like
properties \cite{Curtright-01} of the $\ast$-product:
\begin{eqnarray}
\label{cyclic}
\int dx dp \, A(x,p) \ast B(x,p) 
&=& \int dx dp\, A(x,p) B(x,p) \nonumber \\
&=&\int dx dp\, B(x,p) \ast A(x,p)\,.
\end{eqnarray}
Thus, we have constructed a consistent way of moving back and forth from
quantum operators --acting on a Hilbert space, to complex functions
--acting on the phase space
\begin{eqnarray}
\label{both}
\mathscr{W}(\hat A \hat B) &=& \mathscr{W}(\hat A)
\ast\mathscr{W}(\hat
B)\,,\\ \mathscr{W}^{-1}(A \ast B) &=& \mathscr{W}^{-1}(A)\,\,
\mathscr{W}^{-1}(B)\,.
\end{eqnarray}
This novel product is in agreement with the product rules for the
operators in quantum mechanics. It is non-commutative
\begin{equation}
A \ast B \neq B \ast A\,,
\end{equation}
associative 
\begin{equation}
A \ast (B \ast C) = (A \ast B) \ast C\,,
\end{equation}
and hermitian
\begin{equation}
\overline{A \ast B} = \overline{B} \ast \overline{A}\,,
\end{equation}
where the bar denotes the complex conjugation.

For example, using Bopp's shifts (\ref{bopp}), we obtain:
\begin{eqnarray}
  x \ast p &\equiv& \left( x + \frac{i\hbar}{2}
\overrightarrow{\partial_p}\right)\,p = \ x\, p + \frac{i\hbar}{2} \\
  p \ast x &\equiv& \left( p - \frac{i\hbar}{2}
\overrightarrow{\partial_x}\right)\,x = \ p\, x - \frac{i\hbar}{2}\,.
\end{eqnarray}
Then, the $\ast$-commutator, or the \emph{Moyal bracket} of $x$ and
$p$ is
\begin{equation}
[\,x,\,p\,]_{\ast} \equiv x \ast p - p \ast x = i\hbar \,,
\end{equation}
which is consistent with the canonical commutation relation
$\left[\,\hat x, \,\hat p \,\right] = i \hbar$.  We see that the Moyal
bracket provides a homomorphism with commutators of operators.

\subsection{Moyal bracket and the correspondence principle}
Here we will briefly show that it is the Moyal bracket which provides
a consistent way to quantization, and not the Poisson bracket, as
conjectured by Dirac.

The $\ast$-product and the Moyal bracket are $\hbar$
deformations \cite{Berezin-80,Hirshfeld-02,Hancock-04} of the usual commutative (point-wise) product of
functions $A\,B$, and of the \emph{Poisson bracket} $
\{A\,,\,B\}_P = \partial_x A\, \partial_p B 
- \partial_p A\,\partial_x B$. We can write
\begin{eqnarray}
A \ast B &\!\!\!=\!\!\!&  A\, B + \mathcal{O}(\hbar) \label{h}\,, \\
\frac{1}{i\hbar}[A, B]_{\ast}  &\!\!\!= \!\!\!&\{A, B\}_P + \mathcal{O}(\hbar^2)\,. \label{moy}
\end{eqnarray}
Thus, the name {\em deformation quantization}. Property (\ref{moy})
implies that
\begin{equation}
\lim_{\hbar \to 0} \frac{1}{i\hbar}[A, B]_{\ast} = \{A,B\}_P\,,
\end{equation}
showing that in the classical limit one recovers the Poisson bracket. 

The Moyal bracket obeys Jacobi's identity
\begin{equation}
[[A,B]_{\ast},C]_{\ast} + 
[[C,A]_{\ast},B]_{\ast} +
[[B,C]_{\ast},A]_{\ast} = 0\,,
\end{equation}
as well as the Leibniz rule
\begin{equation}
[A, B \ast C]_{\ast} = [A,B]_{\ast}\ast C + B \ast [A, C]_{\ast}\,.
\end{equation}
Thus, one can endow the space of Weyl's symbols not only with a Lie
algebra structure with respect to the Moyal bracket, but also with an
inner derivative. An analogous statement holds true for the commutator
algebra of quantum operators, but not for the Poisson-bracket algebra.
Groenewold and van Hove showed \cite{Groenewold-46,Hove-51} that there is no
invertible linear map from all functions of phase space $A(x,p),
B(x,p), \ldots$ to Hermitian operators in Hilbert space $\hat A, \hat
B, \ldots$ that will preserve the Poisson bracket structure,
i.e. $\mathscr{W}^{-1} (\{A,B\}_P)=\frac{1}{i
  \hbar}[\mathscr{W}^{-1}(A), \mathscr{W}^{-1}(B)]$. Using Poisson's
bracket as a starting point for quantization works only in the special
cases where the series (\ref{h}) terminates after the first order in
$\hbar$. This is true for functions at most quadratic in $p$ and
$x$. As counterexample, Groenewold \cite{Groenewold-46} showed that the
identically zero expression in Poisson brackets
\begin{equation}
\left\{x^3, p^3\right\}_P + 3\left\{x\,p^2 ,x^2 p \right\}_P = 0\,,
\end{equation}
in Dirac's quantization heuristics, becomes
\begin{equation}
\frac{1}{i\hbar}\left[\hat x^3 , \hat p^3\right] 
+ \frac{3}{i\hbar} 
   \left[ \frac{ \hat x \hat p^2 + \hat p^2 \hat x}{2}\, , \, 
   \frac{\hat x^2  \hat p + \hat p \hat x^2 }{2}\right] = -3\, \hbar^2\,,
\end{equation}
thus exhibiting  a deficiency of order $\hbar^2$.

Through the Weyl's invertible correspondence map (\ref{weyl}), the
Hilbert space of Hermitian operators endowed with the operator
commutator, has as counterpart the algebra of Weyl's symbols endowed
with the Moyal's bracket. Up to an isomorphism, the Lie algebra
generated by the Moyal's bracket is the unique associative
one-parameter deformation of the Poisson bracket \cite
{Vey-75,Flato-76,Bayen-78,Arveson-83,Wilde-83,Gozzi-94}. This uniqueness is extended (up
to an isomorphism) to the $\ast$-product. Thus, it is the Moyal's
bracket that gives the correct correspondence principle for the
quantization scheme \cite{Zachos-05,Groenewold-46,Fronsdal-78}.

\subsection{The Wigner function}
Now that we have a coherent way of modeling quantum operators and
their products, we turn our attention to modeling the quantum state. A
pure state is described by the ket vector $| \psi \rangle$, or
equivalently, by the corresponding density operator $\hat \rho = |
\psi \rangle \langle \psi|$. We will consider here only pure states, because
the generalization to mixed states is straightforward. Let us
denote by $P(x,p)$ the normalized Weyl's symbol of the density
operator $\hat \rho$. We have
\begin{eqnarray}
\label{wigner}
P (x,p) &\equiv& \frac{1}{2\pi\hbar}\,\mathscr{W}(\hat\rho) \\
&=&
	\frac{1}{2\pi}\int dy\, e^{-ipy} \langle x+\frac{\hbar y}{2}\,|\psi \rangle \langle \psi|\, x- \frac{\hbar y}{2}\rangle\,, \nonumber
\end{eqnarray}
or
\begin{equation}
P (x,p) =	
\frac{1}{2\pi}\int dy\, e^{-ipy} \,
\psi(x-\frac{\hbar y}{2}) \,\overline{\psi}(x+\frac{\hbar y}{2})~.
\end{equation}
$P(x,p)$ is the celebrated {\em Wigner function} \cite{Wigner-32}, and
will play a central role in the deformation quantization
technique. Here are several essential properties of the Wigner
function \cite{Iafrate-82,Hillery-83}

\begin{itemize}
\item[(i)] $P(x,p)$ is real.

\item[(ii)]
\begin{eqnarray}
&& \int dp\, P(x,p) = |\psi(x)|^2 = \langle x\, |\, \hat \rho\, | x \rangle \label{wig-q}\\
&& \int dx\, P(x,p) = |\psi(p)|^2 = \langle p\, |\, \hat \rho\, | p \rangle \label{wig-p}\\
&& \int dx dp\, P(x,p) = 1\,. \label{1}
\end{eqnarray}

\item[(iii)] If $P_{\psi}(x,p)$ and $P_{\phi}(x,p)$ correspond to to the states
$\psi(x)$ and $\phi(x)$ respectively, then
\begin{equation}
\label{wig2}
\left|\int dx\, \overline{\psi}(x)\phi(x)\right|^2 
= 2\pi \hbar \int\! dx\,dp P_{\psi}(x,p) P_{\phi}(x,p)\,.
\end{equation}
The last property has two interesting consequences. If $\psi(x) =
\phi(x)$ then
\begin{equation}
\int dx\,dp P_{\psi}^{\,2}(x,p) = \frac{1}{2\pi \hbar}\,,
\end{equation}
and, if we choose $\psi(x)$ orthogonal to $\phi(x)$ , we get
\begin{equation}
\label{neg}
\int dx\,dp\,P_{\psi}(x,p) P_{\phi}(x,p) = 0\,.
\end{equation}
Equation (\ref{neg}) implies that $P(x,p)$ cannot be everywhere
positive. Because it can also take negative values, the Wigner
function is also known as a {\em pseudo-distribution}.

\item[(iv)] $P(x,p)$ is the only pseudo-distribution for which 
each Galilei transformation corresponds to the same Galilei transformation of the quantum mechanical wave function \cite{Wigner-79}. I.e., if $\psi(x) \mapsto\psi(x+a)$, then $P(x,p) \mapsto P(x+a,p)$, and if $
\psi(x) \mapsto\exp(ip'x/\hbar)\, \psi(x)$, then $P(x,p) \mapsto P(x,p-p')$.

\item[(v)] If $\psi(x) \mapsto \psi(-x)$, then $P(x,p) \mapsto P(-x,-p) $,
and if $\psi(x) \mapsto \overline{\psi}(x)$ then $P(x,p) \mapsto
P(x,-p)$.

\end{itemize}

To describe the time evolution of the system in the phase space, let
us take the time derivative of (\ref{wigner}), and use Schr\"odinger's
equation together with its conjugate. We get
\begin{eqnarray*}
\frac{\partial}{\partial t}P(x,p) 
&=& \frac{1}{2\pi i \hbar} \int dy\, e^{-ipy}
 \langle x+\frac{\hbar y}{2}\,| (\hat H \hat \rho - \hat \rho \hat H)|\, x- \frac{\hbar y}{2}\rangle \\
&=& \frac{1}{i\hbar} \mathscr{W}(\left[\hat H, \frac{\hat \rho}{2\pi\hbar}\right])\,,
\end{eqnarray*}
or
\begin{equation}
\label{wigner-ev}
i\hbar \frac{\partial}{\partial t}P(x,p) = [H(x,p),P(x,p)]_{\ast}\,,
\end{equation}
where $H(x,p)$ is the Weyl's symbol of the Hamiltonian $\hat H$.
Equation (\ref{wigner-ev}) is mirroring the time evolution of the
density operator $i \hbar\, \frac{\partial}{\partial t}{\hat \rho} =
[\hat H,\hat \rho]$. The time evolution of the Wigner function can
also be symbolically written using the sine notation
\cite{Bartlett-49}, which emphasizes the non-linear deformation involved by the
Moyal bracket
\begin{equation*}
\frac{\partial}{\partial t}P(x,p) =  
\frac{2}{\hbar}
 \sin\left\{
 \frac{\hbar}{2}\left(
 \frac{\partial}{\partial x}\frac{\partial}{\partial p_1}
 -\frac{\partial}{\partial x_1}\frac{\partial}{\partial p}
 \right)
 \right\} 
 H(x,p)\,P(x_1,p_1) 
\end{equation*}
calculated at $(x_1,p_1)=(x,p)$.

For stationary states $\frac{\partial}{\partial t}P(x,p)=0$, hence the
Hamiltonian \mbox{$\ast$}-commutes with the Wigner function
\begin{equation}
[H(x,p),P(x,p)]_{\ast}=0\,.
\end{equation}
In the stationary case one can get more constraints on the Wigner function. It we apply the Weyl's transform on $\hat H \hat\rho = E\hat\rho$, we obtain the $\ast$-eignevalue equation \cite{Fairlie-64,Curtright-98}
\begin{equation}
\label{star-e}
H(x,p) \ast P(x,p) = E\,P(x,p)\,,
\end{equation}
where $E$ is the energy of the system.  Hermiticiy implies that $\hat
\rho\, \hat H = E \hat \rho$, hence the symmetrical relation holds
true as well
\begin{equation}
P(x,p) \ast H(x,p) = E\, P(x,p)\,.
\end{equation}
Note that if $P_E$ and $P_{E'}$ correspond to the eigenenergies $E$
and $E'$, then due to the associativity of the $\ast$-product, we
have
\begin{equation}
P_E \ast H \ast P_{E'} = E\, P_E \ast P_{E'} = E'\, P_E \ast P_{E'}\,.
\end{equation}
If $E \neq E'$, it follows that $P_E \ast P_{E'}=0$. In the general case we have $ P_E
\ast P_{E'} = \mathscr{W}(\hat \rho_E\, \hat \rho_{E'})/(2\pi\hbar)^2$
which yields the \emph{orthogonality--idempotence} relation of Wigner
functions
\begin{equation}
\label{ortho}
P_E \ast P_{E'} =  \frac{\delta_{E,E'}}{2\pi\hbar}\, P_E\,.
\end{equation}
In deformation quantization, $P(x,p)$ plays an analogous role to the
probability density function in classical statistical mechanics. Namely,
the average of $\hat A$ in state $|\psi\>$ is given by \cite{Moyal-49}
\begin{equation}
\langle \psi| \hat A |\psi \rangle \equiv \langle A(x,p) \rangle 
= \int dx\,dp\, P(x,p)\ast A(x,p)\,.
\end{equation}
Note the resemblance with statistical mechanics. However there is a
major difference in the case of the deformation quantization
formalism: the function $P(x,p)$ is not a probability distribution in
the statistical sense, because it can take negative values.

\subsection{The harmonic oscillator}
To exemplify the concepts introduced so far, let us consider the case
of the simple harmonic oscillator. Without loss of generality we take
$2m=1, \omega=2$. Then the Hamiltonian operator of the harmonic oscillator reads
\begin{equation}
\label{sho}
\hat H =  \hat p^2 + \hat x^2 \,,
\end{equation}
having the corresponding Weyl symbol 
\begin{equation}
\label{sho-w}
 H(x,p) =  p^2 +  x^2\,.
\end{equation}
Let us write the $\ast$-eigenvalue equation (\ref{star-e}) for this
case. We obtain
\begin{equation}
	\left(  p^2 +  x^2 \right) \ast P(x,p) = E\, P(x,p)\,,
\end{equation}
which, using Bopp's shifts representation (\ref{bopp}) becomes
\begin{equation}
\left[
	\left(
	p - \frac{i\hbar}{2}\overrightarrow{\partial_x}
	\right)^2
	+
	\left(
	x + \frac{i\hbar}{2}\overrightarrow{\partial_p}
	\right)^2
\right] P(x,p) = E P(x,p)\,.
\end{equation}
Performing the algebra we arrive to
\begin{equation*}
\left[\,p^2 + x^2 + i\hbar (x \partial_p - p\partial_x) 
- \frac{\hbar^2}{4} (\partial_p^2 + \partial_x^2) -E\, \right] P(x,p) = 0\,.
\end{equation*}
After separating the imaginary and the real parts, we get
\begin{eqnarray}
&&(x \partial_p - p\partial_x)\,P(x,p)=0 \,, \label{im} \\
&& \left[\,p^2 + x^2 - \frac{\hbar^2}{4} (\partial_p^2 + \partial_x^2) 
   -E\, \right] P(x,p) = 0\,. \label{re}
\end{eqnarray}
The symmetry of equation (\ref{im}) indicates that $P(x,p)$ depends
effectively of only one variable, which is a symmetric combination of
$x$ and $p$. Thus, in (\ref{re}) we make the change of variables $t =
2 (p^2 + x^2)/\hbar$ and write $P(t) = e^{-t/2}\,L(t)$. We get
\begin{equation}
\label{laguerre}
\left[\,t\, \partial_t^2 + (1 - t)\, \partial_t + \left(\frac{E}{2\hbar} - \frac{1}{2} \right) \right]L(t) = 0\,.
\end{equation}
For $L(t)$ to be normalizable, the zero derivative term
of (\ref{laguerre}) must be a positive integer
\begin{equation}
\label{en}
\left(\frac{E}{2\hbar} - \frac{1}{2} \right) = n~,\quad n=0,1,2, \dots
\end{equation}
This assures that the series solution of $L(t)$ terminates at a given
rank, hence $L(t)$ is finite and $e^{-t/2}\,L(t)$ is normalizable. But
condition (\ref{en}) yields exactly the quantization formula for the
energy of the harmonic oscillator ($\omega=2$)
\begin{equation}
E_n = 2\hbar\left(n + \frac{1}{2} \right)~,\quad n=0,1,2, \dots
\end{equation}
Consequently, equation (\ref{laguerre}) subject to the constraint
(\ref{en}), becomes the differential equation of Laguerre polynomials
$L_n(t)$ \cite{Chow-00}. Thus, we obtain the analytic expression for the Wigner
functions $P_n(x,p)$ of the harmonic oscillator
\begin{equation}
\label{wigner-ho}
P_n(x,p)=\frac{(-1)^n}{\pi}\,e^{-\frac{p^2 + x^2}{\hbar}}\,L_n\left(
\frac{p^2 + x^2}{\hbar/2} \right)~,~~ n=0,1,2, \dots
\end{equation}
where $L_n$ is the $n$-th Laguerre polynomial. 
\begin{figure}[htb]
\includegraphics[width=\columnwidth]{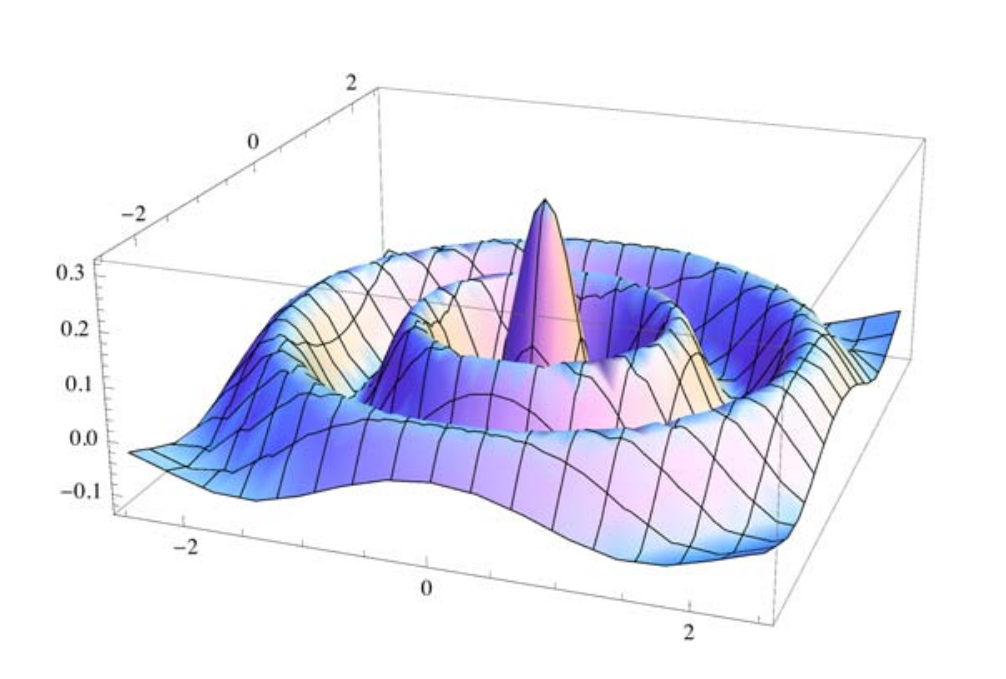}
\caption{The Wigner function $P_5(x,p)$ for the simple harmonic oscillator.}
\label{wigner-5}
\end{figure}
The harmonic oscillator
is one of the few systems where the Wigner functions are completely
known analytically \cite{Akhundova-82}.
In figure (\ref{wigner-5}) we illustrate the Wigner function for $n=5, m=1/2$, and $\omega=2$. 
Note the circular $x \leftrightarrow p$ symmetry of the solution, as
reflected by the equation (\ref{im}). 

Finally, let's analyze the algebraic solution induced by the
factorization of the Hamiltonian in terms of creation and annihilation
operators $\hat a^+ $ and $\hat a$. This method will naturally segue
into the factorization of a general Hamiltonian, which is the crux of
the SUSYQM techniques.  We write the Hamiltonian as
\begin{equation}
\hat H = 2\hbar\left(\hat a^+\, \hat a + 1/2\right)\,,
\end{equation}
where $\hat a = (i\,\hat p + \hat x)/\sqrt{2\hbar}$ and $\hat a^+
\equiv (\hat a)^{\dagger} = (-i\,\hat p + \hat x)/\sqrt{2\hbar}
$. Then $[\hat a\,,\,\hat a^+] = 1$, and after algebraic manipulations
we obtain
\begin{eqnarray}
\hat H |n\rangle &=& 2\hbar\, (n + 1/2)\, |n\rangle~,~~ n=0,1,2,\ldots \\
\hat a^+\,|n\rangle &=& \sqrt{n+1}\, |n+1\rangle \\
\hat a\,|n\rangle &=& \sqrt{n}\, |n-1\rangle \,.
\end{eqnarray}
The harmonic oscillator wave functions are obtained by projecting the
recursion relations
\begin{equation}
|n\rangle = \frac{(\hat a^+)^n}{\sqrt{n!}}\,|0\rangle\,,
\end{equation}
onto the $|x\rangle$ basis, where the starting point is given by $\hat
a\,|0\rangle = 0$.

In the phase space picture these relations become
\begin{equation}
H(x,p) = p^2 + x^2 = 2\hbar\, (a^+ \ast a + 1/2)\,,
\end{equation}
where the annihilation and creation functions $a(x,p)$ and $a^+(x,p)$ are
\begin{equation}
a = (i\,p+x)/\sqrt{2\hbar} \quad , \quad 
a^+ \equiv \overline{( a)} = (-i\,p+x)/\sqrt{2\hbar}\,.
\end{equation}
A simple calculation shows that their Moyal bracket gives
\begin{equation}
[a\,,\,a^+]_{\ast} = 1\,.
\end{equation}
Then, the $\ast$-eignevalue problem becomes
\begin{equation}
H(x,p) \ast P_n(x,p) = 2\hbar\,(n+1/2)\, P_n(x,p)\,,
\end{equation}
where the Wigner function $P_n(x,p)$ is determined from the recursion
relations \cite{Curtright-01}
\begin{equation}
P_n(x,p) = \frac{1}{n!}\,(a^+\ast)^n \, P_0(x,p) \, (\ast\, a)^n\,,
\end{equation}
using as starting point 
\begin{equation}
a \ast P_0(x,p) = 0\,.
\end{equation}
A direct calculation \cite{Curtright-01} shows that the algebraic method recovers the known form 
for the Wigner function (\ref{wigner-ho}). This algebraic technique will be 
extended to a general Hamiltonian, by applying SUSYQM concepts to the phase 
space formalism.

\section{Supersymmetric quantum mechanics}

In this section we give a brief
introduction to supersymmetric quantum mechanics (SUSYQM) techniques which we will use mostly as reference for the next section. For more
details see \cite {Witten-81,Gozzi-83,Cooper-95,Gangopadhyaya-11}. SUSYQM provides an
elegant and useful prescription for obtaining closed analytic
expressions both for the energy eigenvalues and eigenfunctions for a
large class of one dimensional potentials.

Given the Hamiltonian $\hat H = \frac{\hat p^2}{2m} + V(\hat x)$ with
the eigenenergies $E_n$, let us shift its energy spectrum by $E_0$, such
that the new ground state energy becomes zero, and denote the new
Hamiltonian by $\hat H_-$, and its eigenenergies by $E^-_n$:
\begin{equation}
\hat H_- = \frac{\hat p^2}{2m} + V_-(\hat x)\,; \quad 
\hat H_- |n;-\> = E^-_n\, |n;-\>\,, \quad E^-_0 = 0\,.
\end{equation}
$\hat H_-$ can be factorized \cite{Schroedinger-40} as
\begin{equation}
\hat H_- = \hat A^+ \hat A\,,
\end{equation}
where
\begin{equation}
\hat A = \frac{i \hat p}{\sqrt{2m}} + W(\hat x)\,, \quad
\hat A^+ \equiv (\hat A)^{\dagger}= \frac{-i \hat p}{\sqrt{2m}} + W(\hat x)\,.
\end{equation}
The function $\hat W$ is known as the \emph{superpotential} and is related to $V_-(\hat x)$ by 
\begin{equation}
\label{w}
V_-(\hat x)=W^2(\hat x)- \frac{\hbar}{\sqrt{2m}} W'(\hat x)\,.
\end{equation}
Equation (\ref{w}) becomes a Riccati equation when projected on the
$|x\>$ basis, thus allowing for finding $W(x)$. A
sufficient condition for $E^-_0=0$ is given by
\begin{equation}
\label{a}
\hat A\, |\,0;-\> = 0\,,
\end{equation}
which yields the explicit action of $W(\hat x)$ on the ground state
eigenvector $|\,0;-\>$
\begin{equation}
\label{w0}
W(\hat x)\, |\,0;-\> = -\frac{i\hat p}{\sqrt{2m}}\, |\,0;-\> \,.
\end{equation}
If we project (\ref{w0}) on the $|x\>$ basis, we can express $W(x)$ in
terms of the ground state eigenfunction  $\psi_0^-(x)=\< x|\,0;-\>$ as \cite{Gozzi-83}
\begin{equation}
W(x) = -\frac{\hbar}{\sqrt{2m}} 
		\frac{\psi_0^-{}^{\,\prime}(x)}{\psi_0^-(x)}\,.
\end{equation}
By interchanging the operators $\hat A^+$ and $\hat A$ we generate a
new Hamiltonian $\hat H_+ = \hat A\,\hat A^+$, which corresponds to a
new potential $V_+(\hat x)$
\begin{equation}
\hat H_+  = \frac{\hat p^2}{2m} + V_+(\hat x)\,; \quad 
V_+(\hat x) \equiv W^2(\hat x) + \frac{\hbar}{\sqrt{2m}} W'(\hat x)\,.
\end{equation}
Let us denote by $E_n^+$ its eigenenergies: $\hat H_+ |n;+\> =
E_n^+\,|n;+\>$. The two Hamiltonians $\hat H_-$ and $\hat H_+$ are
known as \emph{supersymmetric partner Hamiltonians}. Their
eigenvectors and eigenenergies are related
\begin{eqnarray}
 E_n^{+}  &=& E_{n+1}^{-}\,;\quad E_0^{-} = 0 \,, \label{iso}\\
 |n;+\>   &=& \frac{1}{\sqrt{E_{n+1}^{-}}}\, \hat A\,|n+1;-\> \,, 
 \label{n+} \\
 |n+1;-\> &=& \frac{1}{\sqrt{E_{n}^{+}}}\, \hat A^+\,|n;+\> \,. \label{n-}
\end{eqnarray}
Thus, with the exception of the ground state, the supersymmetric
partner potentials $\hat H_-$ and $\hat H_+$ share the same energy
spectrum (isospectrality), and have interconnected eigenvector
sets. Formulas (\ref {iso} - \ref{n-}) together with (\ref{a})
represent a quick algebraic way of completely finding the spectrum and
eigenvectors of $ \hat H_+$ if we know the spectrum and eigenvectors
of $\hat H_-$. In this way, if $\hat H_-$ corresponds to a very simple
simple system, we have an elegant solution for solving more
complicated cases associated to $\hat H_+$.

There is a special class of partner Hamiltonians for which 
potentials $V_-(\hat x)$ and $V_+(\hat x)$ obey an additional
constraint known as \emph{shape invariance}.  Shape invariance states
that partner potentials are similar in shape 
\begin{equation}
\label{sip}
V_+(\hat x;a_0) = V_-(\hat x;a_1) + R(a_0)\,,
\end{equation}
differing only by a set
of parameters (modeling the \emph{strength} of the interaction) $a_0$, $a_1 = f(a_0)$, and a
constant term $R(a_0)$, independent of $\hat x$. Shape
invariance is a solvability condition that allows in principle to
completely solve the system, as it will be shown shortly.
For this, let us rewrite the shape invariance condition as
\begin{equation}
\label{sih}
H_+(\hat x,\hat p;a_0) + g(a_0) = H_-(\hat x,\hat p;a_1) + g(a_1)\,, 
\end{equation}
where $R(a_0) = g(a_1)-g(a_0)$. Note that we do not have to know the spectrum of one partner Hamiltonian
to know the other. Since Hamiltonians in (\ref{sih}) differ by a
constant, their eigenvalues differ by the same constant, and, up to a
global phase factor, both have the same set of eigenvectors.
Therefore, for all values of $n$ we have
\begin{eqnarray}
E_n^+(a_0) &=& E_n^-(a_1) + g(a_1) - g(a_0)\,, \label{sie}\\
|n,a_0;+\> &=& |n,a_1;-\> \,.
\end{eqnarray}
For normalizable ground states (unbroken SUSY), the
ground state energy of $H_-(\hat x,\hat p;a_i)$ is zero, $E^-_0(a_i)=0$, 
for each
iteration of the parameter $a_i= f(a_{i-1})$. By
successively using the shape invariance (\ref{sie}) and the isospectrality (\ref{iso}) of the
partner Hamiltonians, in conjunction with the unbroken supersymmetry
condition, we get
\begin{eqnarray}
E_0^-(a_0) &=& 0\,, \nonumber\\
E_1^-(a_0) &=& E_0^+(a_0) ~~\mathrm{(isospectrality)} \nonumber\\ 
           &=& E_0^-(a_1)+ g(a_1) - g(a_0) ~~\mathrm{(shape\ invariance)}
           \nonumber\\ 
           &=& g(a_1) - g(a_0) ~~\mathrm{(unbroken\ supersymmetry)}
           \nonumber\\
           &\cdots& \nonumber\\
E_n^-(a_0) &=& g(a_n) - g(a_0)\,.   \label{si}
\end{eqnarray}
Similarly, 
\begin{equation}
\label{siv}
|n,a_0;-\> = \frac{\hat A^+(a_0)\cdots \hat A^+(a_{n-1})\,|\,0,a_n;-\>}
{\left[\prod_{j=0}^{n-1} E^-_{n-j}(a_j)\right]^{1/2}}\,.
\end{equation}
All eigenenergies and eigenvectors of the Hamiltonian $\hat H_-$ can
be determined iteratively by this algorithm. Thus, SUSYQM and shape
invariance determine the entire spectrum of the system without any
need to solve complicated differential equations. For more details see \cite{Gangopadhyaya-11} and the references therein. 

We are now ready to apply these ideas to the phase space
formalism.

\section{SUSYQM and shape invariance in phase space}

SUSYQM on the phase space is obtained by applying Weyl's map to the
operatorial framework discussed above.  Thus, the Weyl symbols of the
partner supersymmetric Hamiltonians $\hat H_-$ and $\hat H_+$ are
\begin{eqnarray}
H_-(x,p) &=& \frac{p^2}{2m} + V_-(x) \equiv A^+(x,p) \ast A(x,p)\,, 
\label{hms}\\
H_+(x,p) &=& \frac{p^2}{2m} + V_+(x) \equiv A(x,p) \ast A^+(x,p)\,,\label{hps}
\end{eqnarray} 
where the annihilation and creation functions are given by
\begin{eqnarray}
A(x,p) &=& \frac{ip}{2m} + W(x)\,; \label{a-star}\\
A^+(x,p) &\equiv& \overline{A}(x,p) = - \frac{ip}{2m} + W(x)\,.\label{a-star1}
\end{eqnarray}
The supersymmetric partner potentials are simply
\begin{equation}
V_{\mp}(x) = W^2(x) \mp \frac{\hbar}{\sqrt{2m}}\,W'(x)\,.
\end{equation}
The last formula can be obtained either by Weyl's mapping the quantum
operators $V_{\mp}(\hat x)$, or by direct calculations from
(\ref{hms}) and (\ref{hps}) using the definitions (\ref{a-star}, \ref{a-star1}), and
the property of Moyal's bracket $[W(x),p]_{\ast} = i\hbar\,\partial_x\,W(x)$.

The $\ast$-eigenvalue problem for the supersymmetric partner
potentials $H_-(x,p)$ and $H_+(x,p)$ written in terms of their
corresponding Wigner functions reads
\begin{eqnarray}
H_-(x,p) \ast P^-_n(x,p) &=& E^-_n\,P^-_n(x,p)\,,\\
H_+(x,p) \ast P^+_n(x,p) &=& E^+_n\,P^+_n(x,p)\,.
\end{eqnarray}
where 
\begin{equation}
P^{\mp}_n(x,p) = \frac{1}{2\pi}\int dy e^{-ipy}
   \< x + \frac{\hbar y}{2} | n ; {\mp}\> \<n ;{\mp}|x - \frac{\hbar y}{2} \>~
\end{equation}
are the $n$-th excited state Wigner functions of the supersymmetric
partner Hamiltonians $H_{\mp}(x,p)$.  The necessary condition for the
ground state energy $E^-_0$ of $H_-(x,p)$ to be zero, is
\begin{equation}
\label{p0}
A(x,p) \ast P^-_0(x,p) = 0\,.
\end{equation}
This yields the $\ast$-product equation
\begin{equation}
W(x)*P^-_0(x,p) = -\frac{i}{\sqrt{2m}} \, p\ast P^-_0(x,p)\,,
\end{equation}
which in terms of Bopp's shifts becomes
\begin{equation}
\label{wp0}
W\left(x+\frac{i\hbar}{2}\, \overrightarrow{\partial_p}\right)\,P^-_0(x,p) 
= \frac{-i}{\sqrt{2m}}
\left( 
   p - \frac{i\hbar}{2}\overrightarrow{\partial_x}
\right) P^-_0(x,p)\,.
\end{equation}

By applying the Weyl map on equations (\ref{n+}) and (\ref{n-}) we
obtain \cite{Curtright-98} the connections between the $\ast$-eigenfunctions
of the supersymmetric partner potentials $H_-(x,p)$ and $H_+(x,p)$
\begin{eqnarray}
P^+_n(x,p) &\!\!\!=\!\!\!& \frac{1}{E^-_{n+1}}A(x,p) \ast P^-_{n+1}(x,p)\ast A^+(x,p) \label{ppn}\,, \\
P^-_{n+1}(x,p) &\!\!\!=\!\!\!& \frac{1}{E^+_n} \,A^+(x,p) \ast P^+_n(x,p) \ast A(x,p) \label{pnp}\,.
\end{eqnarray}

The only constraint we imposed so far was that the Hamiltonians
$H_{\mp}(x,p)$ are supersymmetric partners. This leads, besides the
isospectrality, to the $\ast$-product connection between the
corresponding Wigner functions $P^{\mp}_n(x,p)$. It is almost the
``carbon copy '' of the SUSYQM case. Next, we explore the implications
of the shape invariance. In phase space, equation (\ref{sih}) becomes
\begin{equation}
\label{sih-star}
H_+(x,p;a_0) + g(a_0) = H_-(x,p;a_1) + g(a_1)\,,
\end{equation}
where $a_0$ and $a_1 = f(a_0)$. Because the Hamiltonians in (\ref{sih-star}) differ
by a constant, they have the same set of $\ast$-eigenfunctions. Hence,
up to a sign factor (because Wigner functions are
real), shape invariance implies
\begin{equation}
\label{si-p}
P^+_n(x,p;a_0) = P^-_n(x,p;a_1)\,,
\end{equation}
and the energy spectrum 
\begin{equation}
E^-_n (a_0) = g(a_n) - g(a_0)~,\quad \mathrm{with\ } a_n =f^n(a_0).
\end{equation}
In addition, the shape invariance  together with the
supersymmetric condition, leads to a new recursion formula among the
Wigner functions:
\begin{equation}
\label{wigner-si}
P^-_n(x,p;a_0) = A^+(x,p;a_0) \ast \frac{P^-_{n-1}(x,p;a_1)}{E^-_n(a_0)} \ast A(x,p;a_0)
\,.
\end{equation}
The proof is immediate:
\begin{eqnarray*}
P^-_n(x,p;a_0) &=& A^+(x,p;a_0) \ast \frac{P^+_{n-1}(x,p;a_0)}{E^+_{n-1}(a_0)} \ast A(x,p;a_0)\,,\\
   &=& A^+(x,p;a_0) \ast \frac{P^-_{n-1}(x,p;a_1)}{E^-_{n}(a_0)} \ast A(x,p;a_0)
   \,, 
\end{eqnarray*}
where the first equality follows from SUSYQM, and the second one from shape invariance and isospectrality. We can now iterate (\ref{wigner-si}) to obtain the expression for a
general $P^-_n(x,p;a_0)$ starting from the ground state Wigner function
$P^-_0(x,p;a_n)$. We have
\begin{eqnarray}
\label{wig-si}
P^-_n(x,p;a_0) &=& 
A^+(x,p;a_0) \ast \cdots \ast A^+(x,p;a_{n-1}) \nonumber \\
&& \ast \frac{P^-_0(x,p;a_n)}{\prod_{j=0}^{n-1} E^-_{n-j}(a_j)} \ast A(x,p;a_{n-1}) \nonumber\\
&& \ast \cdots \ast A(x,p;a_0)\,.
\end{eqnarray}
This recursion formula together with (\ref{p0}) as its starting point,
determines, the entire set of Wigner functions for shape invariant
systems.  However, for concrete examples, the multiple $\ast$-products
can become prohibitively difficult to calculate. The harmonic oscillator
is one a the few cases where we can determine analytically the entire
sequence of Wigner functions.

\subsection{The Morse potential}

As an example, let us consider the one dimensional Morse potential. The Morse oscillator is a
good approximation of the oscillatory motion of the bi-atomic
molecules \cite{Lee-82,Belchev-10}. Without loss of generality, we take
$\hbar = 1$ and $2m=1$.  Then, the Hamiltonian reads
\begin{equation}
H_-(x,p;a) = p^2 + V_-(x,a)\,,
\end{equation}
where the Morse potential
\begin{equation}
V_-(x,a) = a^2 + b^2 e^{-2sx} - 2b (a+1/2)e^{-sx}~;\quad  a,b,s > 0\,,
\end{equation}
can be generated as $V_- = W^2 - W'$ from the superpotential \cite{Cooper-95}
\begin{equation}
W(x;a)= a - b e^{-sx}\,.
\end{equation}
The corresponding annihilation and creation phase space functions are
\begin{equation}
A(x,p) =  ip + a - b e^{-sx}~; \quad A^+(x,p) = - ip + a - b e^{-sx}\,.
\end{equation}
As we know \cite{Cooper-95}, Morse potential is shape invariant. The relevant
parameters of the model are $a_0 = a$ and $a_1 \equiv f(a_0) = a - s$
respectively. Thus
\begin{equation}
H_+(x,p;a) + g(a) = H_-(x,p;a-s) + g(a-s)~; \quad g(a) = a^2\,,
\end{equation}
and consequently, the energy spectrum is 
\begin{equation}
E^-_n(a) = a^2 - (a-ns)^2\,.
\end{equation}
Let's find the ground state Wigner function $P^-_0(x,p)$ using
\begin{equation}
\label{anih}
A(x,p) \ast P^-_0(x,p) = 0\,.
\end{equation}
Writing the $\ast$-products in (\ref{anih}) as Bopp's shifts, we obtain
\begin{equation}
\left\{
i
\left( 
   p - \frac{i}{2}\overrightarrow{\partial_x} 
\right) + a- b e^{
   - s\left( x+\frac{i}{2}\, \overrightarrow{\partial_p}\right) 
   }
\right\} \,P^-_0(x,p) 
= 0 \,,
\end{equation}
which becomes a mixed differential/finite-difference equation
\begin{eqnarray}
\label{brutal-1}
(a + ip)\,P^-_0(x,p) &=& b \, e^{-sx} P^-_0(x,p-is/2) \nonumber \\
 &-& \frac{1}{2}\partial_x P^-_0(x,p)\,.
\end{eqnarray}
The above equation mixes the partial derivative with respect to one
variable with a shift with respect to the other variable. This hints
to a well-known \cite{Lebedev-72} differential/finite-difference formula for the
modified Bessel functions $K_{\nu}(x)$ 
\begin{equation}
2\, \partial_x\,K_{\nu}(x) + K_{{\nu}-1}(x) + K_{1+{\nu}}(x) = 0\,.
\end{equation}
The exponential term in the r.h.s. of equation (\ref{brutal-1})
suggests that $K_{\nu}(x)$ depends on $x$ via an exponential term. The
shift in momentum variable suggests that $p$ appears in the index ${\nu}$
of $K_{\nu}(x)$.  Therefore we substitute in (\ref{brutal-1}) the
following ansatz
\begin{equation}
\label{smart}
P^-_0(x,p) = C_1 e^{-C_2\, x} K_{\alpha(p)}\left(C_3\, e^{-sx}\right)\,,
\end{equation}
where the constants $C_1,C_2,C_3$, and the function $\alpha(p)$ are to
be determined. Plugging (\ref{smart}) into (\ref{brutal-1}), after
some algebra, we find that (\ref{brutal-1}) is identically satisfied
for
\begin{equation}
\label{ai} 
C_2 = 2a~,~~
C_3 = \frac{2b}{s}~;~~\mathrm{and\ }~
\alpha(p) = \frac{2ip}{s}\,.
\end{equation}
for any value of $C_1$, which is constrained by the
normalization of the Wigner function (\ref{1}). We obtain
$C_1=\frac{2}{\pi s}(2b/a)^{2a/s}$. Thus, the ground state Wigner
function for the Morse oscillator is
\begin{equation}
\label{morse}
P^-_0(x,p;a) = \frac{2}{\pi s}\left(\frac{2b}{a}\right)^{2a/s}
   e^{-2a x} K_{2ip/s}\left(\frac{2b}{s}\, e^{-sx}\right)\,.
\end{equation}

\begin{figure}[htb]
\includegraphics*[width=\columnwidth]{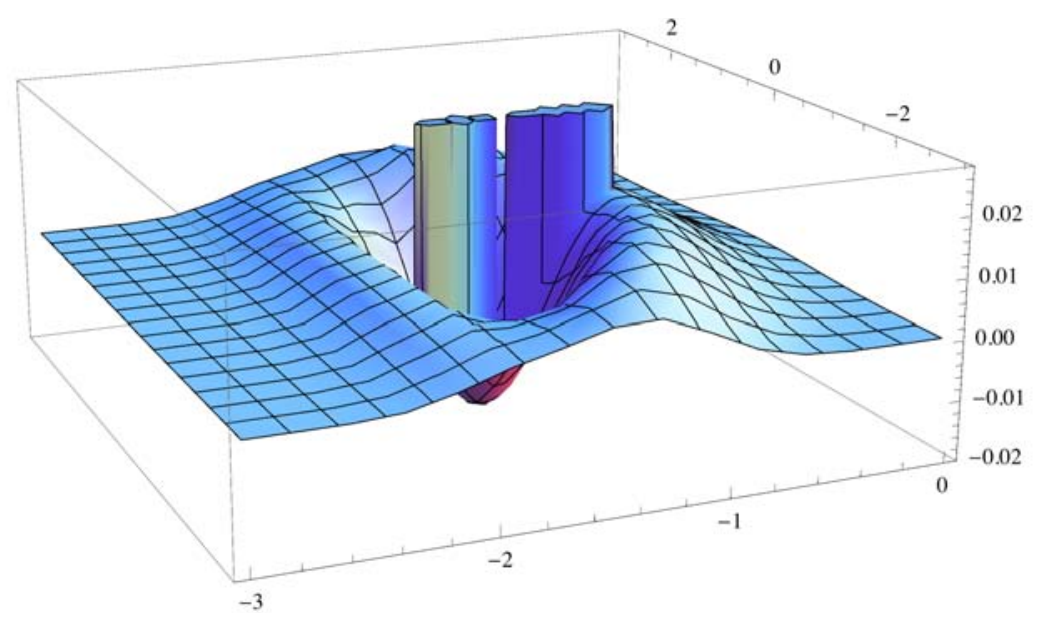}
\caption{The Wigner function $P_1(x,p)$ for the Morse oscillator, for
  $a=5,b=1$, and $s=1$.}
\label{mor-wig1}
\end{figure}

Finding the next excited states Wigner functions is now straightforward using the shape invariance (\ref{wigner-si}). For example, let us find $P^-_1(x,p;a)$ 
\begin{eqnarray*}
\label{p1}
&&P^-_1(x,p;a) = A^+(x,p;a) \ast \frac{P^-_0(x,p;a-s)}{E^-_1(a_0)} \ast A(x,p;a) \\
&&=\left[a -ip - b\,e^{-sx} \right] \ast \frac{P^-_0(x,p;a-s)}{a^2-(a-s)^2}\ast \left[a+ip - b\,e^{-sx} \right] \\
&&=\left[a -i\left(\!p-\frac{i}{2}\overrightarrow{\partial_x}\!\right) 
- b\,e^{-s(x+\frac{i}{2}\overrightarrow{\partial_p})} \right] \!
\frac{P^-_0(x,p;a-s)}{2as -s^2} \\
&&\quad\left[a +i\left(\!p+\frac{i}{2}\overleftarrow{\partial_x}\!\right) 
- b\,e^{-s(x-\frac{i}{2}\overleftarrow{\partial_p})} \right]\,. 
\end{eqnarray*}
After some algebra, we arrive to
\begin{equation}
P^-_1(x,p;a) = \alpha(x)
\,K_{\nu}(y) - \beta(x) \left[
K_{\nu-1}(y) + K_{\nu+1}(y) \right]\,,
\end{equation}
where $\nu=\frac{2ip}{s}; y=\frac{2b}{s}\,e^{-sx}$, and
\begin{eqnarray*}
\alpha(x)&=&\frac{2^{\frac{2 a}{s}} e^{-2 a x} \left(\frac{b}{a-s}\right)^{-2+\frac{2 a}{s}} }{ 2\pi  (2 a-s) s^2}\left[4 b^2+e^{2 s x} (2 a-s)^2\right]\,,\\
\beta(x)&=&\frac{4^{a/s} b e^{(-2 a+s) x} \left(\frac{b}{a-s}\right)^{-2+\frac{2 a}{s}}}{\pi  s^2}\,.
\end{eqnarray*}
In figure (\ref{mor-wig1}) we illustrate the Wigner function $P_1(x,p)$ for the Morse
oscillator. The graph is cut on the vertical axis to
emphasize the details near the origin.

\section{Conclusions}

We have shown that quantum mechanics can be expressed equivalently in the 
language of ``normal'' functions defined on the phase space, endowed with the $
\ast$-product. The Moyal bracket corresponds to the operator commutator, and one 
can use it as an heuristic tool for quantization. In the classical limit ($\hbar 
\to 0$) the Moyal bracket becomes simply the Poisson bracket, and the $\ast$-
product, becomes the normal commutative product of functions. In this picture, 
quantum mechanics appears as a deformation of the classical mechanics, with the 
deformation parameter $\hbar$.  Supersymmetric quantum mechanics induces 
additional relations between the Wigner functions, while shape invariance exposes 
a simple approach to recursively obtain the Wigner function of a system, starting 
from its ground state Wigner function, as exemplified for the Morse potential.

\section*{Acknowledgements}
I would like to acknowledge a sabbatical leave and grant from Columbia College Chicago that made this work possible. I am grateful to Prof. Ennio Gozzi for hospitality extended to me during my visit to the Department of Physics at the University of Trieste -Miramare Campus (Italy), and for fruitful discussions. I would like to thank INFN for financial support (grant GE41), and CERN (Geneva, Switzerland) for hospitality.

\bibliographystyle{plain}
\bibliography{quantum}

\end{document}